\documentclass[%
 reprint,
 superscriptaddress,
 amsmath,amssymb,
 aps,
 pra,
]{revtex4-2}

\usepackage{graphicx}% Include figure files
\usepackage{dcolumn}% Align table columns on decimal point
\usepackage{bm}% bold math
\usepackage{verbatim} % Multiline comment
\usepackage{amsmath}
\usepackage{float}
\usepackage{soul}
\usepackage{xcolor}
\usepackage{soul, color, xcolor}

\usepackage{subfigure}
\begin{document}

%\preprint{APS/123-QED}

\title{Shortcuts to Quantum Approximate Optimization Algorithm}% Force line breaks with \\

\author{Yahui Chai}
\email{cyh@originqc.com}
\affiliation{Origin Quantum Computing Company Limited, Hefei 230026, China}

\author{Yong-Jian Han}
\affiliation{CAS Key Laboratory of Quantum Information (University of Science and Technology of China), Hefei 230026, China}
\author{Yu-Chun Wu}
\affiliation{CAS Key Laboratory of Quantum Information (University of Science and Technology of China), Hefei 230026, China}

\author{Ye Li}
\affiliation{Origin Quantum Computing Company Limited, Hefei 230026, China}

\author{Menghan Dou}
\affiliation{Origin Quantum Computing Company Limited, Hefei 230026, China}

\author{Guo-Ping Guo}
\email{gpguo@ustc.edu.cn}
\affiliation{Origin Quantum Computing Company Limited, Hefei 230026, China}
\affiliation{CAS Key Laboratory of Quantum Information (University of Science and Technology of China), Hefei 230026, China}


\date{\today}% It is always \today, today,

\begin{abstract}
The Quantum Approximate Optimization Algorithm (QAOA) is a quantum-classical hybrid algorithm intending to find the ground state of a target Hamiltonian.  Theoretically, QAOA can obtain the approximate solution if the quantum circuit is deep enough.  Actually, the performance of QAOA decreases practically if the quantum circuit is deep since near-term devices are not noise-free and the errors caused by noise accumulate as the quantum circuit increases.  In order to reduce the depth of quantum circuits, we propose a new ansatz dubbed as ``Shortcuts to QAOA'' (S-QAOA), S-QAOA provides shortcuts to the ground state of target Hamiltonian by including more two-body interactions and releasing the parameter freedoms.  To be specific, besides the existing ZZ interaction in the QAOA ansatz, other two-body interactions are introduced in the S-QAOA ansatz such that the approximate solutions could be obtained with smaller circuit depth.  Considering the MaxCut problem and Sherrington-Kirkpatrick (SK) model, numerically computation shows the YY interaction has the best performance.  The reason for this might arise from the counterdiabatic effect generated by YY interaction. On top of this, we release the freedom of parameters of two-body interactions, which a priori do not necessarily have to be fully identical, and numerical results show that it is worth paying the extra cost of having more parameter freedom since one has a greater improvement on success rate. 
\end{abstract}

\maketitle

%\tableofcontents

\section{Introduction}
In the noisy intermediate-scale quantum (NISQ) era~\cite{nisq_2018}, the number of reliable quantum operations is limited by the quantum errors which contains quantum decoherence, rotation error, and so on. Thus people are interested in the quantum-classical hybrid algorithm whose quantum circuit depth is decreased with the help of classical optimizers, like the Quantum Approximate Optimization Algorithm (QAOA)~\cite{farhi2014quantum} which is expected to get an approximate solution for combinatorial optimization problems. In this hybrid algorithm, the quantum state is prepared by a quantum computer, the parameters in the quantum circuit are optimized by a classical optimizer to find an evolution path that needs less circuit depth. Furthermore, QAOA is expected to have a better performance than quantum adiabatic algorithms (QAA) and get a quantum advantage using near-term devices ~\cite{performance2020,cd_qaoa}. However, the performance of QAOA is limited by the noise of the near-term devices~\cite{xue2019effects}, and Google's experiment for QAOA with their Sycamore superconducting qubit quantum processor shows that errors overwhelm the theoretical performance increase at larger layers~\cite{google2021}. Therefore executing QAOA on near-term quantum computers is a challenging task. It is important to reduce the circuit depth of QAOA to make it achievable for near-term devices.

QAOA can be regarded as a digitized and variational version of QAA. QAA starts with a simple-to-prepared ground state of an initial Hamiltonian, and the adiabatic theorem guarantees that the ground state of the final Hamiltonian can be obtained if the time-dependent Hamiltonian varies slowly. The adiabatic condition requires that the running time of QAA scales as $T \sim O(1/{\Delta_{min}}^2)$, $\Delta_{min}$ is the minimal spectral gap~\cite{qaa2018} of the quantum system during the evolution, thus many problems are hard to optimize using QAA because of the overlong annealing time. Besides, the digitization of QAA may lead to a deep quantum circuit to minimize the trotter error~\cite{Poulin2011,Suzuki:1976be}. QAOA is expected to overcome these problems by optimizing parameters (including evolution path and duration for every digitized step) using a classical optimizer. The optimized parameters of QAOA are related to a fast evolution path, so QAOA is expected to break through the limits of the adiabatic condition~\cite{performance2020}.

Shortcuts to adiabaticity (STA)~\cite{sta2013,sta2019} is a class of methods to accelerate the quantum adiabatic process, counterdiabatic (CD) driving~\cite{cd-driving1,cd-driving2,Berry_2009} is a technique of STA to reduce the finite-speed diabatic effect by adding counter terms to the time-dependent Hamiltonian. The CD driving Hamiltonian has a better performance and reduces the evolution time compared to adiabatic evolution~\cite{dqc_2021}. Besides, Ref.~\cite{Alan_2015} proposes a superadiabatic route to implement universal quantum computation by using CD driving, and energy cost of STA via CD driving is studied at Ref.~\cite{Alan_2015, Alan_2016}. In addition, energy cost optimization of CD driving, through the correct choice of the counterdiabatic Hamiltonian is studied in Ref.~\cite{Alan_2017}. Recently, Ref.~\cite{cd_qaoa} indicates QAOA is at least counterdiabatic and has a better performance than finite time adiabatic evolution. There is also an effort to add a conterdiabatic term to the ansatz of QAOA to reduce the quantum circuit depth~\cite{DCDQAOA2021, zhu2020adaptive}.

Our work focuses on the MaxCut problems and Sherrington-Kirkpatrick (SK) Model, both of them are NP-hard problems, and the QAOA is expected to have a quantum acceleration on these problems. In this work, we investigate the counterdiabatic effect of QAOA, and the simulation result implies that adding a two-gate term associate with the YY interaction to the quantum circuit will accelerate the optimization of QAOA. Besides, we release the QAOA parameter freedom of the two-body interactions (including ZZ interaction and YY interaction) to reduce the quantum circuit depth, eg. each two-body interaction has its independent parameter. The idea of extending the parameter degree is mentioned in 2017~\cite{farhi2017quantum} and they use random initial parameters of each ZZ interaction. In our case, firstly, we optimize the QAOA parameters to get an optimal global value, and then use the optimal parameters of QAOA as the initial parameters of each two-body term to do a further local optimization. The proposed algorithm uses the philosophy of STA, so we call it ``Shortcuts to QAOA'' (S-QAOA), and the simulated result shows that S-QAOA can get a good result with a shallower quantum circuit compared with QAOA.

\section{Problems and Quantum Algorithms}
 In this work, we focus on the MaxCut problem and SK model. The MaxCut problem is defined on a graph $G(V,E)$, $V = \{1,2, \cdots ,n\}$ is the set of vertices, and $E = \{ \left( (i,j), w_{ij} \right) \}$ is the set of edges where $(i,j)$ is a pair of connected vertices and $w_{ij}$ is the weight of the edge $(i,j)$. We study two classes of graphs: the first class is unweighted 3-regular graphs (u3R) whose weights $w_{ij}$ are a constant for the all edges, eg. $w_{ij} = 1, \forall (i,j)\in E$; the second class is weighted 3-regular graphs (w3R) whose weights $w_{ij}$ are uniform random numbers in range $[0,1]$.
The target Hamiltonian of MaxCut problem is:
\begin{equation}
   H_C = -\sum_{(i,j) \in E} w_{ij} \frac{I-Z_i Z_j}{2}.
\end{equation}
The SK model is defined on the complete graph and the coefficient $w_{ij}$ is randomly chosen from the set $\{-1, 1\}$. The Hamiltonian of the SK model is :
\begin{equation}
    H_C = \sum_{i<j} w_{ij} Z_i Z_j.
\end{equation}

\subsection{\label{sec:level1} Quantum Adiabatic Algorithms}
QAA is able to find the ground state of a target Hamiltonian if the annealing process is slow enough, and its feasibility is guaranteed by the adiabatic theorem. In QAA, a simple-to-prepared quantum state is chosen as the initial state, which usually is the uniform superposition over computational basis states: $\psi(t=0)=|+\rangle^{\otimes n}$, $n$ is the number of qubits of the quantum system.  $\psi(t=0)$ is the ground state of Hamiltonian $H_B = -\sum_{i=1}^n X_i$. The time-dependent Hamiltonian of QAA is:
\begin{equation}
\begin{aligned}
    H(t) = (1-\lambda(t)) H_B + \lambda(t) H_C, \\
    0 \leq t \leq T, \lambda(0) = 0, \lambda(T) = 1,
\end{aligned}
\end{equation}
where $H_C$ represents the target Hamiltonian. The time evolved state driven by $H$ would be:
\begin{equation}
    |\psi(T)\rangle = \mathcal{T} e^{-i\int_{0}^T dt H(t)} |\psi(0)\rangle,
\end{equation}
where $\hbar=1$ and $\mathcal{T}$ is the time-ordering operator. The approximate ground state of $H_C$ can be obtained in the end $t = T$ if the Hamiltonian $H$ varies adiabatically.

STA is able to suppress the finite-speed diabatic excitations by adding the CD driving term to the Hamiltonian $H$~\cite{Berry_2009}, this CD term is theoretically known and given by  adiabatic
gauge potential (AGP) $A_{\lambda}$~\cite{AGP_2017}:
\begin{equation}\label{H_sta}
    H_{CD}(t) = H + \dot{\lambda} A_{\lambda}.
\end{equation}
However, it requires the knowledge of the spectral properties of the instantaneous Hamiltonian to get the AGP, and the exact AGP is typically nonlocal, which makes the experimental implementation difficult~\cite{cd_experiment}. The exact AGP can be approximated by a local parameterized gauge potential $A_{\lambda}(\boldsymbol{\alpha})$, and the requirement of $A_{\lambda}(\boldsymbol{\alpha})$ is to minimize the action $S(A_{\lambda})$:
\begin{equation}
\begin{aligned}
    \mathop{Min}\limits_{\boldsymbol{\alpha}}& : S\left( A_{\lambda}(\boldsymbol{\alpha}) \right)
    = Tr\left[ G^2(A_{\lambda}) \right], \\
    &G(A_{\lambda}) = \partial_{\lambda}A_{\lambda} + i[A_{\lambda}, H],
\end{aligned}
\end{equation}
the coefficient $\boldsymbol{\alpha}$ can be determined variationally~\cite{variation_AGP_2017}. Furthermore, the approximate AGP can be systematically constructed by the nested commutators~\cite{cd_expansion_2019}:
\begin{equation}
    A_{\lambda}^l(\boldsymbol{\alpha}) = i\sum_{k=1}^l \alpha_k(t) \underbrace{[H,[H,\cdots [H,}_{2k-1} \partial_{\lambda}H]]].
\end{equation}
The first order of the approximate AGP is the commutator of the initial Hamiltonian $H_B$ and the target Hamiltonian $H_C$, so the counterdiabatic Hamiltonian becomes:
\begin{equation}\label{H_CD}
\begin{aligned}
H_{CD}^1 &= H + \dot{\lambda} A_{\lambda}^1 \\
     &=  (1-\lambda(t)) H_B + \lambda(t) H_C + i \dot{\lambda} \alpha_1 [H_B, H_C],
\end{aligned}
\end{equation}
and it can be proven that the coefficient $\alpha_1$ is negative~\cite{cd_qaoa}. In $l= +\infty$ limits, the CD driving terms $A_{\lambda}^l(\boldsymbol{\alpha})$ are able to compensate the diabatic excitations exactly, but it is necessary to just consider the first few orders of $A_{\lambda}^l(\boldsymbol{\alpha})$ to avoid the nonlocal operations.

\subsection{Quantum Approximate Optimization Algorithm}
The ansatz of QAOA  consists of a series of digitized evolution, and the parameters of QAOA are optimized by a classical optimizer~\cite{farhi2014quantum}:
\begin{equation}
    |\psi(\boldsymbol{\beta},\boldsymbol{\gamma})\rangle = e^{-i\beta_p H_B} e^{-i\gamma_p H_C} \cdots e^{-i\beta_1 H_B} e^{-i\gamma_1 H_C} |\psi(0)\rangle.
\end{equation}
The purpose of the classical optimization is to find the optimal parameters to minimize the expectation of the target Hamiltonian:
\begin{equation}
\begin{aligned}
    \mathop{Min}\limits_{\boldsymbol{\beta}, \boldsymbol{\gamma}}& : E(\boldsymbol{\beta},\boldsymbol{\gamma}) =  \langle \psi(\boldsymbol{\beta},\boldsymbol{\gamma})| H_C | \psi(\boldsymbol{\beta},\boldsymbol{\gamma}) \rangle ,\\
    &\boldsymbol{\beta}=(\beta_1,\cdots,\beta_p), \boldsymbol{\gamma}=(\gamma_1,\cdots,\gamma_p).
\end{aligned}
\end{equation}
If the layer $p$ of QAOA is large enough, the expectation $E$ will decrease during the optimization and converge to the target Hamiltonian ground state. The effective Hamiltonian of QAOA can be constructed by the second order Baker-Campbell-Hausdorff (BCH) expansion~\cite{cd_qaoa}:
\begin{equation}\label{H_eff}
\begin{aligned}
& e^{-i\beta_k H_B} e^{-i \gamma_k H_C} \sim e^{-i H_{eff}}, \ 1\leq k \leq p,\\
&H_{eff} = \gamma_k H_C + \beta_k H_B - \frac{i \beta_k \gamma_k}{2} [H_B, H_C]
\end{aligned}
\end{equation}
Compare the result of Eq.\ref{H_CD} and Eq.\ref{H_eff}, we can find that the effective Hamiltonian $H_{eff}$ of QAOA has a first-order CD driving term, and the coefficient of this term is also negative. So the evolution of QAOA will include some counterdiabatic effects to compensate the diabatic excitation.

\subsection{Shortcuts to QAOA}

Inspired by the technique of STA, we consider the counterdiabatic effect of QAOA by introducing more two-body interactions: $H_{M} = \sum\limits_{(i,j)\in E}\frac{P_iQ_j + Q_iP_j}{2}, PQ \in \{YZ, YY,XX,XZ,XY\}$. The ansatz and the effective Hamiltonian become:
\begin{equation}\label{H_eff_sqaoa}
\begin{aligned}
    &e^{-i\alpha_k H_M} e^{-i\beta_k H_B} e^{-i\gamma_k H_C} \sim e^{-i H_{eff}} , \ 1 \leq k \leq p,\\
    &H_{eff} = \alpha H_M + \beta H_B + \gamma H_C \\
    &-i\frac{\beta\gamma}{2}[H_B,H_C] - i\frac{\alpha\beta}{2}[H_M, H_B] - i\frac{\alpha\gamma}{2}[H_M, H_C],
\end{aligned}
\end{equation}
where $\alpha_k, k=1,\cdots,p$ are the variational parameters for $H_M$. The effective Hamiltonian in Eq.\ref{H_eff_sqaoa} might have more CD terms and would accelerate the process of quantum optimization.
Introducing more interactions led to a deeper quantum circuit, this problem can be partially solved by a compact implementation of the $H_M$ and $H_C$ (Eq.\ref{S-QAOA ansatz}), which does not need extra SWAP gates if using a local connectivity superconducting quantum computer. Furthermore, the combination of YY/XX and ZZ interaction can be realized using two CNOT gates and some single-qubit gates, which has the same number of CNOT gates as the way only implementing ZZ interaction (the details of this part will be discussed in Sec.\ref{discussion}). In this case, the circuit depth of S-QAOA and QAOA can be regarded as the same, because the errors caused by CNOT gates are more serious than single-qubit gates.

Another difference between S-QAOA and QAOA is the optimization of the parameters: in S-QAOA, after the optimization of QAOA parameters, the parameter freedoms of two-body interactions are released and a further optimization is performed. More parameters will improve the expressivity of the quantum circuit and can get a better result than QAOA in the same quantum layer. The procedure of S-QAOA is as follows:\\

1. Optimize the QAOA parameters using INTERP strategy~\cite{performance2020}, get the optimal parameters $(\tilde{\boldsymbol{\beta}},\tilde{\boldsymbol{\gamma}})_p$ for layer $p$.\\

2. Release the parameter freedom of $ZZ$ interaction, and introduce an extra two-body interaction $M_{ij} = \frac{P_iQ_j + Q_iP_j}{2}, PQ \in \{YZ, YY,XX,XZ,XY\}$, followed by each $ZZ$ interaction. Operation of each layer $k \in [1,p]$ of S-QAOA is:
\begin{equation}\label{S-QAOA ansatz}
\begin{aligned}
    & e^{i\beta_k \sum\limits_i^n X_i}e^{-i \gamma_k\sum\limits_{ij} w_{ij} Z_i Z_j} \rightarrow \\
    & e^{i\beta_k \sum\limits_i^n X_i} \prod\limits_{ij}\{e^{-i \gamma^{ij}_k w_{ij} Z_i Z_j} e^{-i\alpha_k \gamma^{ij}_k w_{ij} M_{ij}} \}.
\end{aligned}
\end{equation}
The initial parameters of S-QAOA are: $\beta_{k}= \tilde{\beta}_k, \gamma_{k}^{ij}= \tilde{\gamma}_k, \alpha_k = 0$. The strength of $M_{ij}$ interaction should be positively related to that of $Z_iZ_j$ interaction, so the parameter $\gamma_{ij}$ is added to each $M_{ij}$ interaction (This is the ansatz of S-QAOA for SK model, and it is the same for MaxCut problem except for a coeffecient $\frac{1}{2}$. For simplicity, we will only show the ansatz for SK model in the following.)

3. Use the finite-difference method to calculate the gradients of parameters $(\{\gamma^{ij}_k\}, \beta_k, \alpha_k)$:
\begin{equation}
\begin{aligned}
&g(\theta) = \frac{E(\theta + \epsilon) - E(\theta)}{\epsilon}, \\ 
&\forall \theta \in (\{\gamma^{ij}_k\}, \beta_k, \alpha_k), k = 1, \cdots ,p,
\end{aligned}
\end{equation}
$\epsilon$ is a small constant. Set a threshold $\delta_1$ and if $|g(\theta)|>\delta_1$, add $\theta$ to set $A$.\\

4. Optimize the parameters in set $A$ until convergent, if the decrease of energy is smaller than a threshold $\delta_2$ after the optimization, exit; else, update the optimized parameters and return to step3.

\section{Simulation Result}
We study the u3R and w3R MaxCut problem on 14-vertex graphs, and SK model on 6-vertex graphs. In each case, the results are averaged on 20 random graphs. There are three ansatzes being studied: QAOA; only releasing the parameter freedom of ZZ interaction (This ansatz only contains ZZ interaction, and for simplicity, we call it `ZZ technique' below); adding an extra two-body interaction and releasing the freedom of parameters (S-QAOA). The simulation result implies that the third ansatz has the best performance in all cases, and the suitable extra two-gate term is able to accelerate the optimization process significantly. A comprehensive study of the optimal type of extra two-gate term in S-QAOA can be found in Fig.\ref{fig:all_s_exp_poss}, which shows the performance and the comparison of all the possible extra two-gate types: $M_{ij} = \frac{P_iQ_j + Q_iP_j}{2}, PQ \in \{YZ, YY,XX,XZ,XY\}$. In general, $YY$ interaction has the best performance, and it is included in S-QAOA ansatz. Thus the operation of each layer $k \in [1,p]$ of S-QAOA is:
\begin{equation}
    e^{i\beta_k \sum\limits_i^n X_i} \prod_{ij} \{e^{-i \gamma^{ij}_k w_{ij} Z_i Z_j} e^{-i\alpha_k \gamma^{ij}_k w_{ij} Y_i Y_j} \}.
\end{equation}

\begin{figure}[htbp]
	\centering
	\includegraphics[width=8cm]{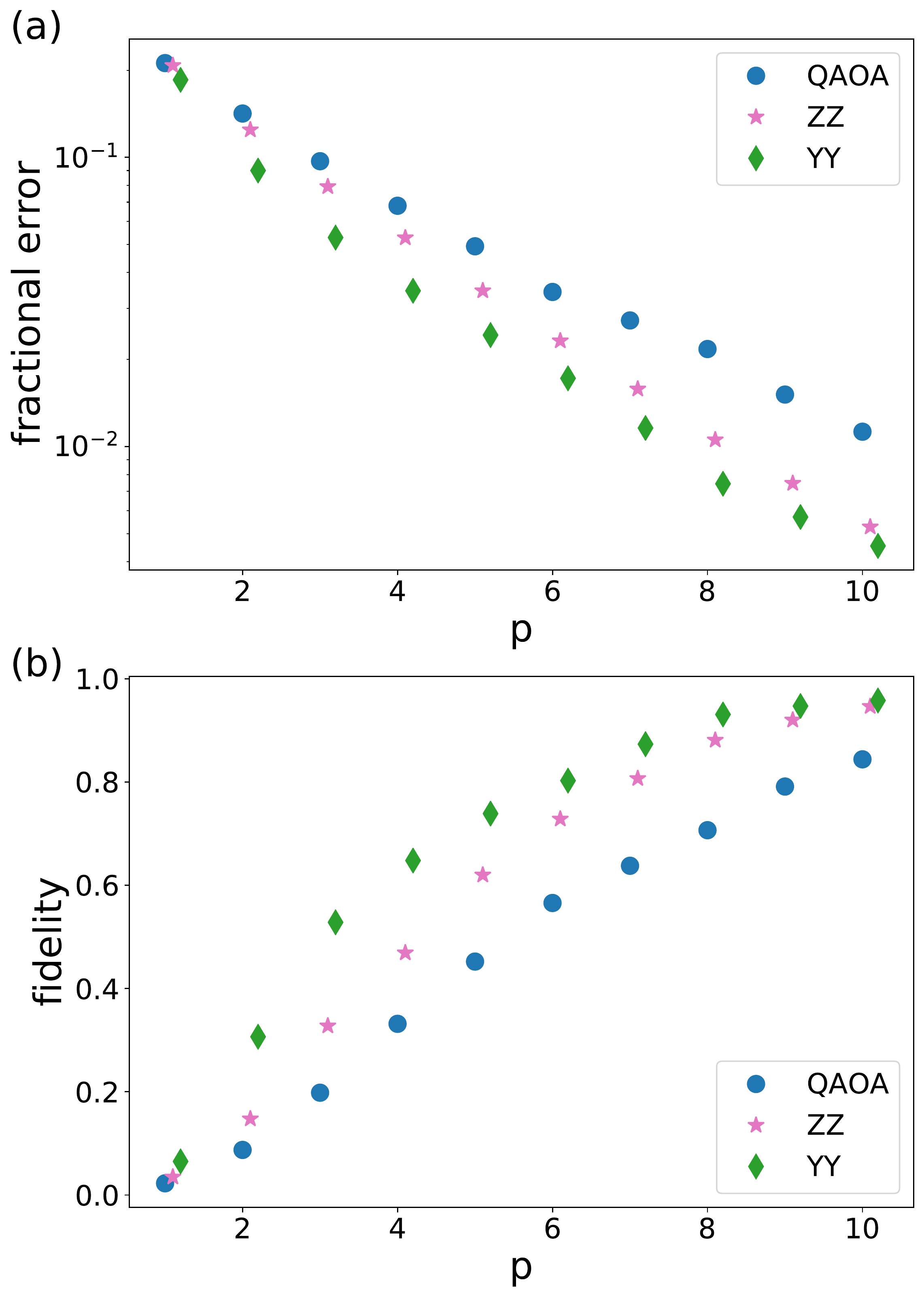}
	\caption{Comparison of the performance of QAOA, $ZZ$ technique and S-QAOA for MaxCut problem on 14-vertex u3R graphs. $(a)$The fractional error of the results obtained by these algorithms, the definition of the fractional error is : $r = 1-\frac{E}{E_{opt}}$, $E_{opt}$ is the theoretical optimal value. $(b)$The fidelity obtained by these algorithms, the fidelity represents the possibility of the optimal solutions in the final state. Label `ZZ' represents the ansatz which only releases the freedom of parameters (ZZ technique). Label `YY' represents the S-QAOA ansatz.} 
	\label{fig:u3r}
\end{figure}
\begin{figure}[htbp]
	\centering
	\includegraphics[width=8cm]{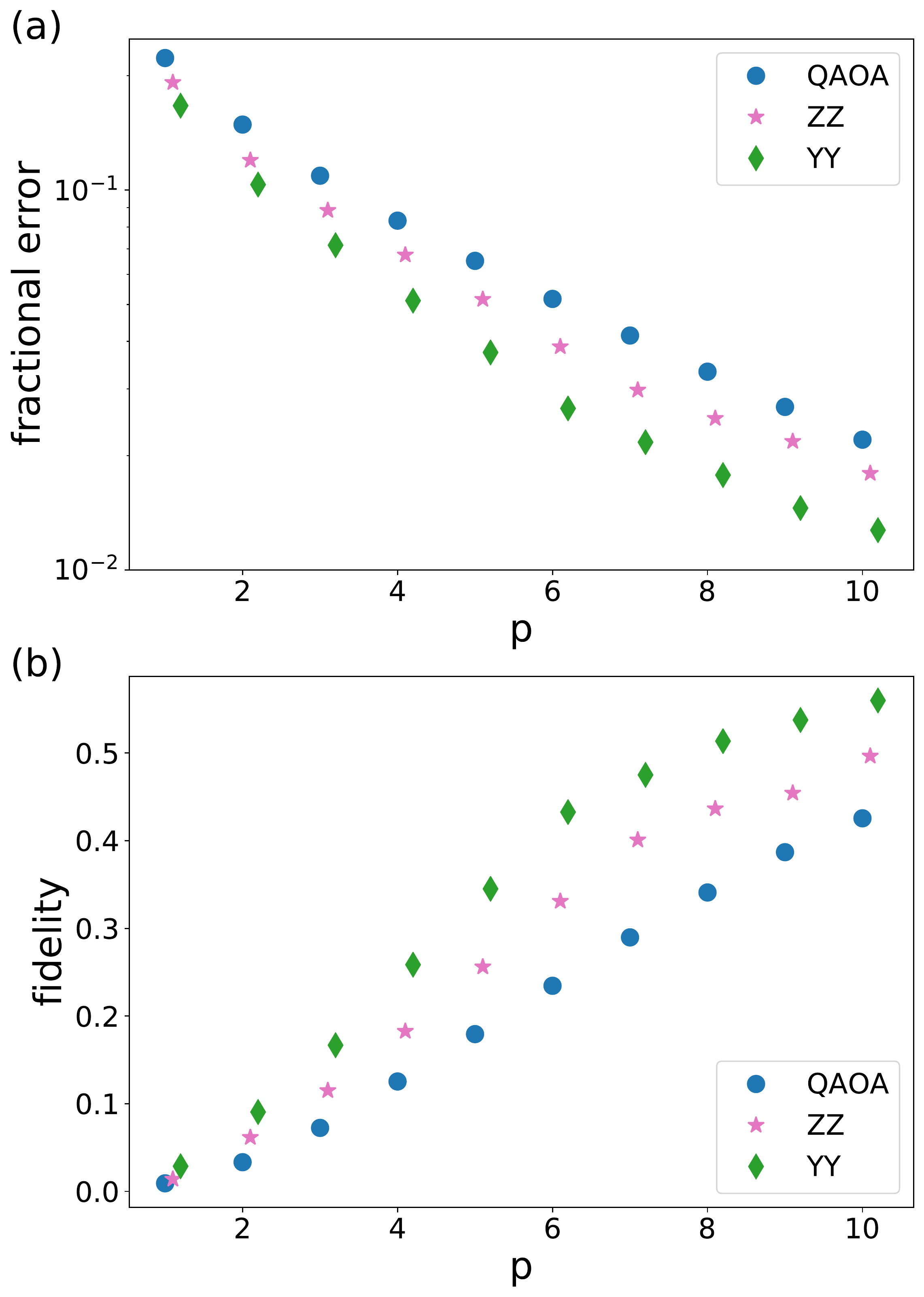}
	\caption{Comparison of (a) the fractional error and (b) the fidelity of QAOA , ZZ technique and S-QAOA for MaxCut problem on 14-vertex w3R graphs. The performance of S-QAOA is obviously better than QAOA at the same layer. The fidelity of S-QAOA in $p=4,5$ is rough twice as much as that of QAOA. Moreover, if the quantum layer $p$ is increased, S-QAOA is still able to improve the fidelity of the result. For example, S-QAOA gets a $38\%$ improvement of the fidelity over QAOA at $p=10$.} 
	\label{fig:w3r}
\end{figure}

The results of u3R MaxCut problem are shown in Fig.\ref{fig:u3r}. Obviously, only releasing the freedom of parameters of the existing $ZZ$ interaction will produce a better result in the same layer than QAOA. On top of this, adding the $YY$ interaction will improve the performance further, especially at the low layer where the performance of QAOA is not so good. When the quantum layer $p$ increases, the difference between the three ansatz is smaller, this is due to that the evolution time is large enough for QAOA if the quantum layer is large and there is little space for improvement. Fig.\ref{fig:w3r} demonstrate the simulation results of w3R graphs, and the superiority of S-QAOA is more obvious in this case, the average fidelity is improved significantly even at $p=10$. The MaxCut problem on w3R graphs is more difficult than u3R graphs, since the energy gap of the MaxCut Hamiltonian on w3R graph is smaller than that of u3R graph, and it needs a longer evolution time to satisfy the adiabatic condition. The result shows the potential of S-QAOA to solve the problems which are difficult for QAA and QAOA. 

\begin{figure}[htbp]
	\centering
	\includegraphics[width=8cm]{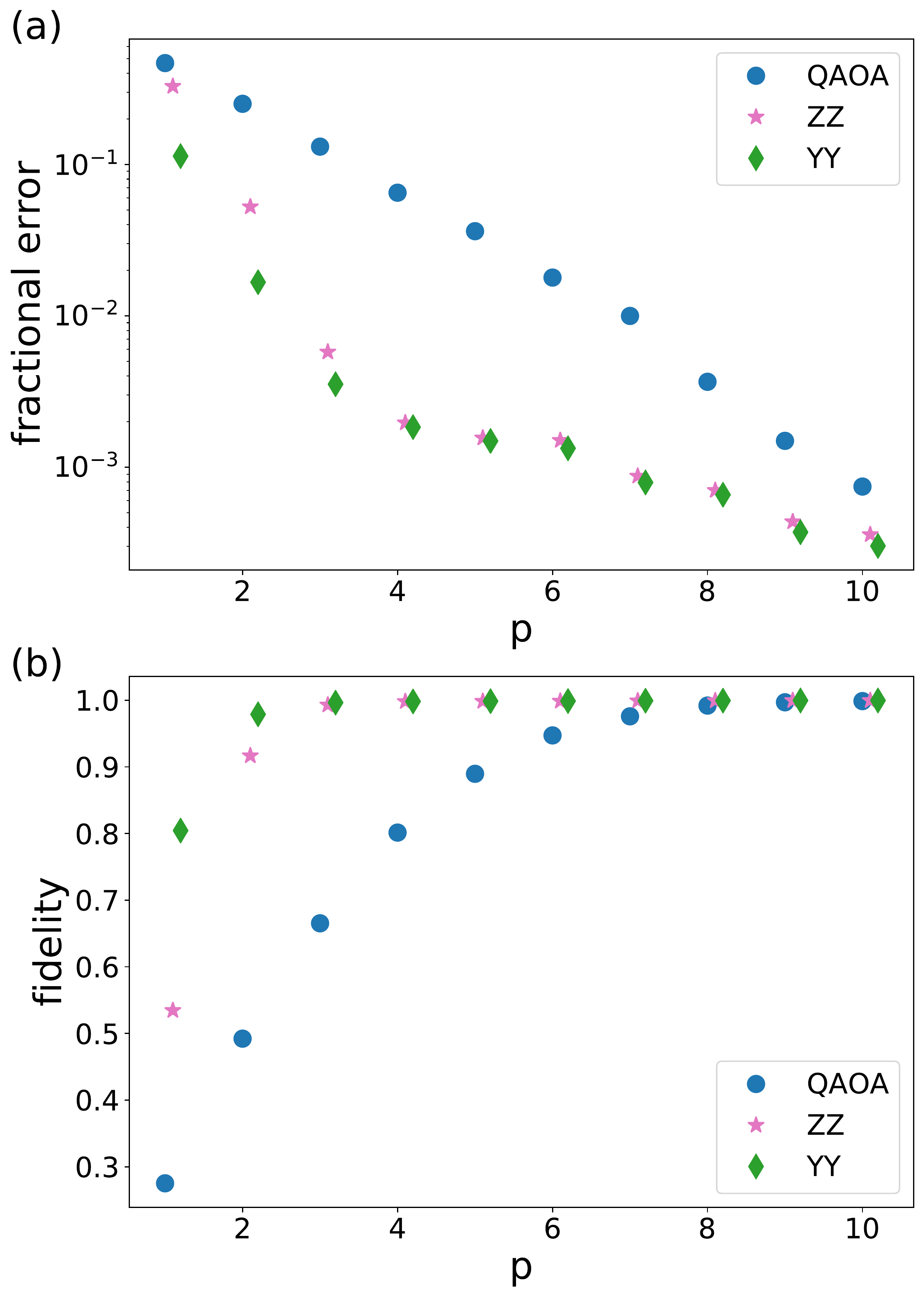}
	\caption{Comparison of (a) the fractional error and (b) the fidelity of QAOA, ZZ technique and S-QAOA for 6-vertex SK model. $ZZ$ technique can significantly decrease the fractional error and improve the fidelity at low layers, and the result can be improved further by using S-QAOA with a few additional parameters.} 
	\label{fig:sk}
\end{figure}

SK model is defined on the complete graph, it is challenging to implement SK model on a NISQ device which has a limited qubit connectivity. Because of the all-to-all interaction of SK model, all the nodes can be entangled together at $p=1$, and there are sufficient parameters to optimize if the parameters of ZZ interaction are independent. So that a pretty good result can be obtained at $p=1$ if only releasing the parameter freedom (Fig.\ref{fig:sk}). Furthermore, if a $YY$ interaction is added to the ansatz, the fidelity is obviously improved and reaches about $80\%$ at $p=1$. S-QAOA introduces only one extra parameter compared with the ZZ technique, and the performance of S-QAOA is significantly better than the latter at $p=1$. The significant improvement produced by $YY$ interaction confirms its effects on countering the diabatic excitations and accelerating the process of quantum optimization. The difference between S-QAOA and $ZZ$ technique becomes smaller and smaller when the quantum layer $p$ is increased, and this is due to that the parameter freedoms of $ZZ$ technique are enough for the optimization at large layer $p$, and there is not much space for S-QAOA to improve.

\begin{figure}
    \centering
    \includegraphics[width=8cm]{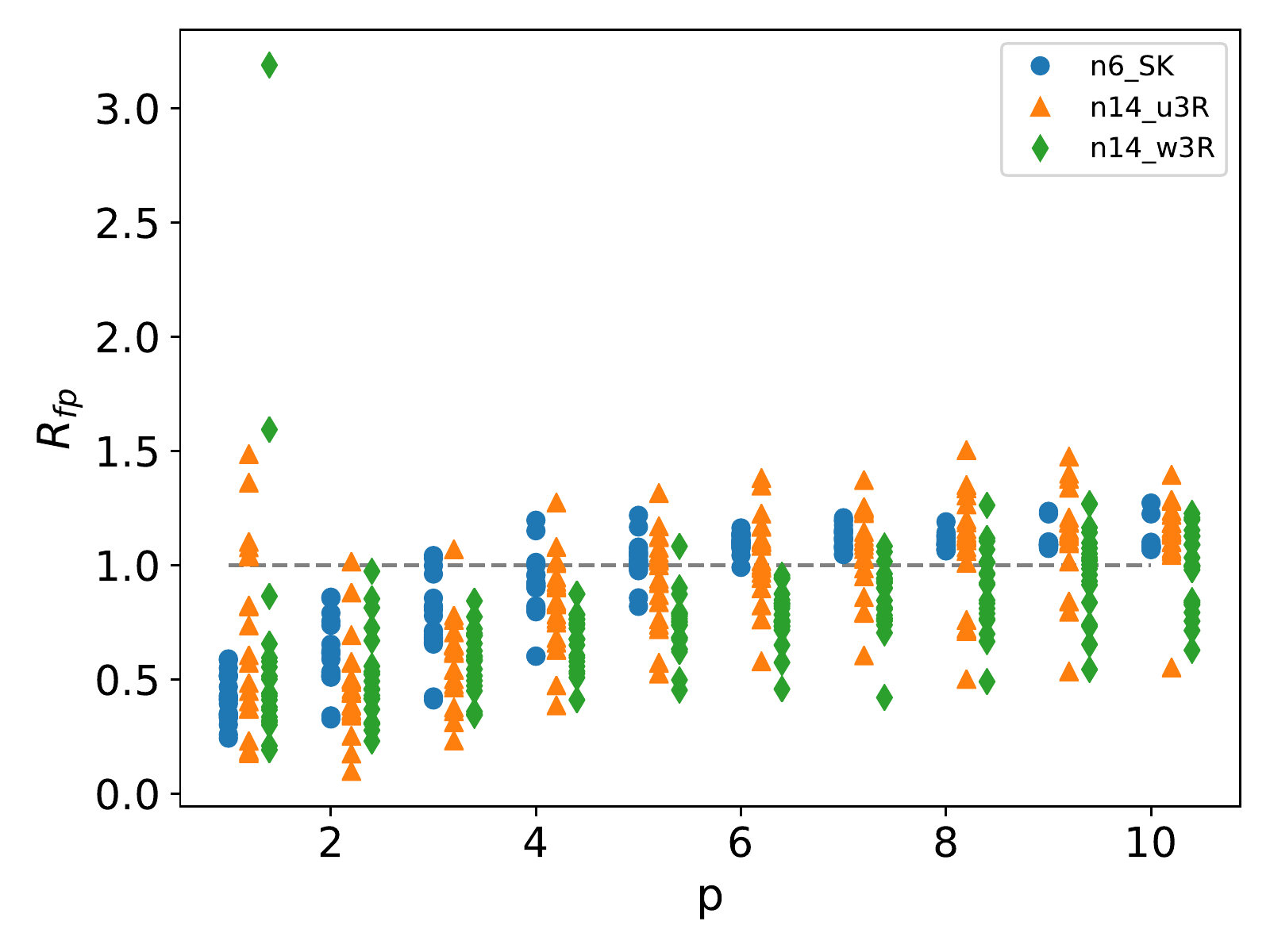}
    \caption{The ratio of function evaluations and fidelity for u3R, w3R MaxCut problems and SK model. We include 20 random instances for each problem. There are some unusual data in w3R MaxCut case, and the reason is that the cost function of S-QAOA is the expectation of the energy, so it is possible to get a lower fidelity with a better expectation value. These unusual points disappear when p increases.}
    \label{fig:R_fp}
\end{figure}

S-QAOA has a better performance in all cases we study, e.g. for a specific quantum layer, $R_p = p_{S}/p_{Q}>1$, $p_{S}(p_{Q})$ is the possibility of the optimal solution got by S-QAOA(QAOA). S-QAOA does a further optimization and has more parameters compared with QAOA, so the number of function evaluations of S-QAOA is more than that of QAOA, e.g. $R_f=f_S/f_Q > 1$, $f_S(f_Q)$ is the cumulative number of function evaluations of S-QAOA(QAOA), which will faithfully reflect the total cost of the algorithm. More specifically, $f_S = f_Q + f_G + f_O$, where $f_G$ is the number of function evaluations for calculating the gradients of parameters, and $f_O$ is the number of function evaluations for further optimizations in S-QAOA.
It is necessary to consider whether it is cost-effective to do these extra optimizations. We show the ratio of $R_f$ and $R_p$: $R_{fp}=R_f / R_p$ in Fig.\ref{fig:R_fp}, and $R_{fp} < 1$ represents that it is deserved to do the extra optimizations of S-QAOA to produce a higher improvement of fidelity. It is clearly that the ratio $R_{fp} < 1$ if $p\leq 4$ for almost cases. If p increases, QAOA can produce quite high fidelity for SK model and u3R MaxCut problem. There is little space to improve for S-QAOA, so the ratio $R_{fp}$ of SK model and u3R MaxCut problem approach to 1 or even larger than 1 for large p. For w3R MaxCut problem, the fidelity of QAOA is far away from 1, so S-QAOA can improve the fidelity effectively with some further optimizations. In all, S-QAOA is an effective way to improve the result of QAOA, especially in case QAOA has limited performance.

\section{Discussion}\label{discussion}
The quantum circuit of QAOA consists of the alternating implementation of $e^{-i\gamma H_C}$ and $e^{-i \beta H_B}$, the nest commutator of $H_B$ and $H_C$ is able to span the entire Lie algebra associated with the Hilbert space of the n-qubit system. QAOA can approximate any element of the entire unitary group $U(2^n)$ if a sufficient deep quantum circuit is applied~\cite{lie_1, lie_2_2020}. Based on the existing generator $H_B$ and $H_C$, S-QAOA provides an extra generator associated with the two-body interaction to accelerate the process of approximating the desired unitary operation. The numerical result shows that the generator associated with the YY interaction has the best performance, and the quantum circuit depth is reduced significantly by including it in S-QAOA ansatz (Fig.\ref{fig:all_s_exp_poss}).
The advantage of YY interaction might be explained by the connection of it and the CD driving terms (Eq.\ref{H_sta}). For MaxCut problem and SK model, the first order of Eq.\ref{H_sta} is:
\begin{equation}
\begin{aligned}
    &H_{CD}^1 = i \alpha_1 [H_B, H_C]\\
    &= -2 \alpha_1 \sum_{(i,j)} w_{ij} (Z_iY_j + Y_iZ_j).
\end{aligned}
\end{equation}

There is a little improvement when we add the above $YZ$ term to the quantum circuit Eq.\ref{S-QAOA ansatz}(Fig.\ref{fig:all_s_exp_poss}). The limited improvement of $YZ$ term is possibly due to that the effective Hamiltonian of QAOA contains the first order of CD driving terms (Eq.\ref{H_eff}). In order to introduce more counter terms to compensate the diabatic excitations, we consider the second order of CD driving terms(Eq.\ref{S-QAOA ansatz}):
\begin{equation}
    H \sim \beta H_B + \gamma H_C,
\end{equation}
\begin{equation}\label{H_cd2}
\begin{aligned}
    &H_{CD}^2 = i \alpha_2 [H,[H,[H_B, H_C]]]\\
    &= -\alpha_2 \{(c_1\beta^2+c_2\gamma^2) \sum_{(i,j)} w_{ij} (Y_i Z_j + Z_i Y_j) \\
    &+ c_3 \beta\gamma \sum_{i}\sum_{jk\in\{n_i\}} w_{ij}w_{ik} (X_i Z_j Y_k + X_i Y_j Z_k)\\
    &+ c_4 \gamma^2 \sum_{i}\sum_{jkm\in \{n_i\}} w_{ij}w_{ik}w_{im} Y_i Z_j Z_k Z_m \},
\end{aligned}
\end{equation}
There are some positive coefficients $c_1, c_2, c_3, c_4$, and $\{n_i\}$ represent the neighbors of vertex $i$, eg. $(i, j) \in E, \forall j \in \{n_i\} $  The first two terms on the right side of equation Eq.\ref{H_cd2} can be generated if a series of $YY$ interaction $H_Y = \sum_{ij} w_{ij} Y_i Y_j$  is added to ansatz:
\begin{equation}
\begin{aligned}
    &e^{-i\alpha H_Y} e^{-i\beta H_B} e^{-i\gamma H_C} \sim e^{-i H_{eff}}, \\
    &H_{eff} = \alpha H_Y + \beta H_B + \gamma H_C \\
    &-i\frac{\beta\gamma}{2}[H_B,H_C] - i\frac{\alpha\beta}{2}[H_Y, H_B] - i\frac{\alpha\gamma}{2}[H_Y, H_C]\\
    &=  \alpha H_Y +\beta H_B + \gamma H_C + H_1 + H_2 + H_3.
\end{aligned}
\end{equation}
$H_1$ has the same form as the first order of CD driving terms (Eq.\ref{H_CD}). $H_2$ and $H_3$ will generate terms same as the first two terms on the right side of equation Eq.\ref{H_cd2}:
\begin{equation}
\begin{aligned}
    &H_2 = - i\frac{\alpha\beta}{2}[H_Y, H_B] = \alpha\beta \sum_{ij} w_{ij} (Y_i Z_j + Z_i Y_j).
\end{aligned}
\end{equation}

\begin{equation}
\begin{aligned}
    &H_3 = - i\frac{\alpha\gamma}{2}[H_Y, H_C] \\
    &=\alpha\gamma \sum_{i}\sum_{jk\in \{n_i\}} w_{ij}w_{ik} (X_i Y_j Z_k + X_i Z_j Y_k).
\end{aligned}
\end{equation}

$H_2$ contains the same interaction as the first term on the right side of equation Eq.\ref{H_cd2}, and $H_3$ contains the same interaction as the second term. Besides, the sign of the coefficient of $H_2$ and $H_3$ are the same, it is the same for that of Eq.\ref{H_cd2}. So it can partially compensate the excited states by adding the $YY$ interaction to S-QAOA ansatz.

Besides, the parameter freedom of ZZ interaction is released to further reduce the quantum circuit depth, and since the strength of $YY$ and $ZZ$ interactions should be positively related, the coefficient $\gamma_{ij}$ is added to each YY interaction:
\begin{equation}\label{yy_not_gather}
\begin{aligned}
    &e^{-i\alpha H_Y} e^{-i\beta H_B}  e^{-i\gamma H_C} \rightarrow \\
    &e^{-i \alpha \sum\limits_{(ij)}w_{ij}  \gamma_{ij} Y_i Y_j} e^{-i\beta\sum\limits_{i=1}^n X_i}\  e^{-i \sum\limits_{(ij)} w_{ij}  \gamma_{ij} Z_i Z_j}.
\end{aligned}
\end{equation}
The coefficient $\alpha$ plays the same role as in STA, and is also be determined variationally.

\begin{figure}[h]
    \centering
    \includegraphics[width=8cm]{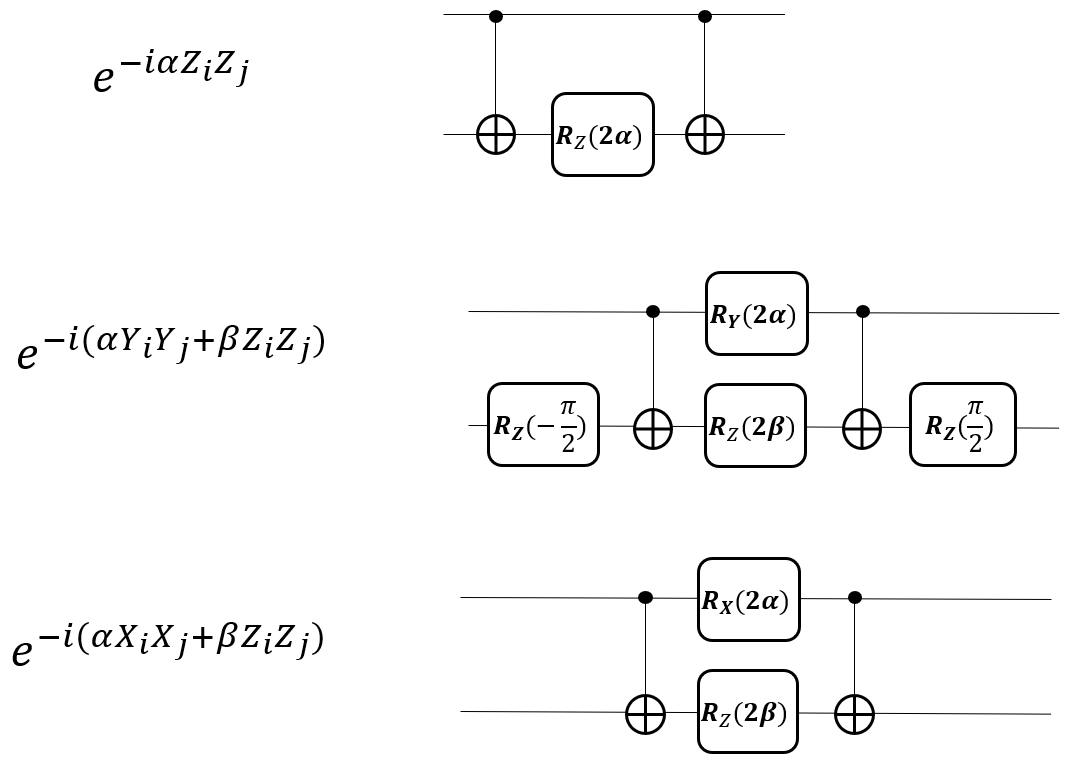}
    %\caption{Caption}
    \caption{The implementations of the two-body interactions in QAOA and S-QAOA using CNOT gate. The definitions of the single-qubit rotation gates are : $R_X(\theta) = e^{-i\frac{\theta}{2}X}, R_Y(\theta) = e^{-i\frac{\theta}{2}Y}, R_Z(\theta) = e^{-i\frac{\theta}{2}Z}$. As shown above, the number of CNOT gate will be the same for QAOA and S-QAOA if introducing YY or XX interaction to S-QAOA. Other two-body interaction types can be implemented using three CNOT gates as shown at Ref.~\cite{two_qubit_gate_2004}.}
    \label{fig:circuit}
\end{figure}

\begin{figure}[h]
    \centering
    \includegraphics[width=8.5cm]{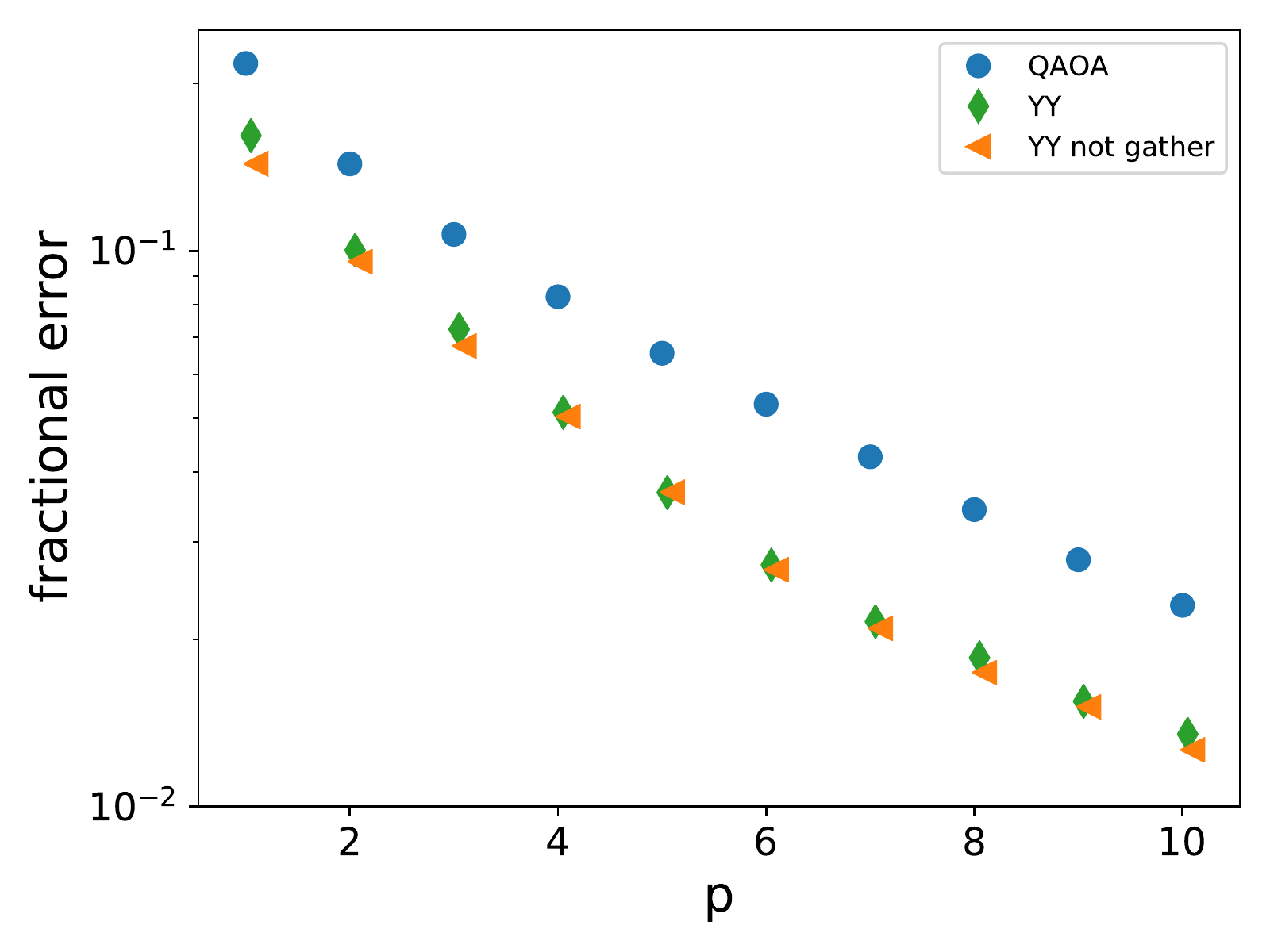}
    %\caption{Caption}
    \caption{The average fractional error of MaxCut problem on 10 random w3R graphs for the comparison of Eq.\ref{yy_not_gather} and Eq.\ref{yy_gather}. Label `YY not gather' represents the ansatz in Eq.\ref{yy_not_gather}. Label `YY' represents the ansatz in Eq.\ref{yy_gather}, e.g., combining the YY and ZZ interactions together.}
    \label{fig:yy_order}
\end{figure}

To reduce the number of CNOT gates, the order of operations in the ansatz is adjusted, and the operations of the $YY$ and $ZZ$ interactions are combined together:
\begin{equation}\label{yy_gather}
    e^{-i\beta\sum\limits_i X_i}  \prod\limits_{(ij)} e^{-i w_{ij} \gamma_{ij}(\alpha Y_i Y_j + Z_i Z_j)}.
\end{equation}
The combined implementation of YY and ZZ interactions can be realized as shown in Fig.\ref{fig:circuit}. Fig.\ref{fig:yy_order} shows the comparison of the performance of ansatz in Eq.\ref{yy_not_gather} and Eq.\ref{yy_gather} by considering the MaxCut problem on w3R graphs. The result implies that the performance of those two ansatzes are basically the same, so we choose the ansatz in Eq.\ref{yy_gather} to reduce the quantum circuit depth.

\section{Summary and outlook}
In this work, S-QAOA ansatz is proposed to reduce the required quantum circuit depth. The main innovation of S-QAOA is: firstly, the extra two-body interaction is considered in the S-QAOA to compensate the diabatic effect and accelerate the process of quantum optimization; secondly, the parameter freedoms of two-body interactions are released to enhance the capacity of the quantum circuit. We study the performance of S-QAOA and QAOA on MaxCut problem and SK model, and the simulation result implies that S-QAOA has better performance at lower quantum layers compared with QAOA. So S-QAOA is a good candidate to solve the combinatorial problems using NISQ.

Releasing the parameter freedom needs extra cost of optimization, the numerical simulation shows it is cost-effective because of the greater improvement on fidelity. In S-QAOA, further optimization is performed on the parameters that have large gradients, and there needs more work to explore how to release the parameter freedoms properly, e.g., how to pick out the most critical parameters to do a further optimization. The most influential parameters should be different for different problems, and it is important to develop an efficient way to pick them out. Besides, in our primary exploration, introducing more parameters in S-QAOA makes the optimization more challenging when considering the shot noise, so the optimization methods that are more robust to noise should be considered in further work, like COBYLA, SPSA, etc. Furthermore, it deserves further exploration to explain the reason for the YY interaction's effectiveness more clearly. We will study more cases and do simulations with noise to test and improve our idea in the next step.

\begin{acknowledgments}
We thank Lingxiao Xu, Cheng Xue, Huanyu Liu and Qingsong Li for valuable discussions and suggestions. The results of this work are simulated using pyQpanda, and the pyQpanda package can be downloaded at https://github.com/OriginQ/QPanda-2.
\end{acknowledgments}

\newpage
\bibliography{sqaoa}% Produces the bibliography via BibTeX.

\appendix
\newpage
\pagebreak
\widetext

\section{Results With Error Bars}

\begin{figure*}[htbp]
	\centering
	\includegraphics[width=18cm]{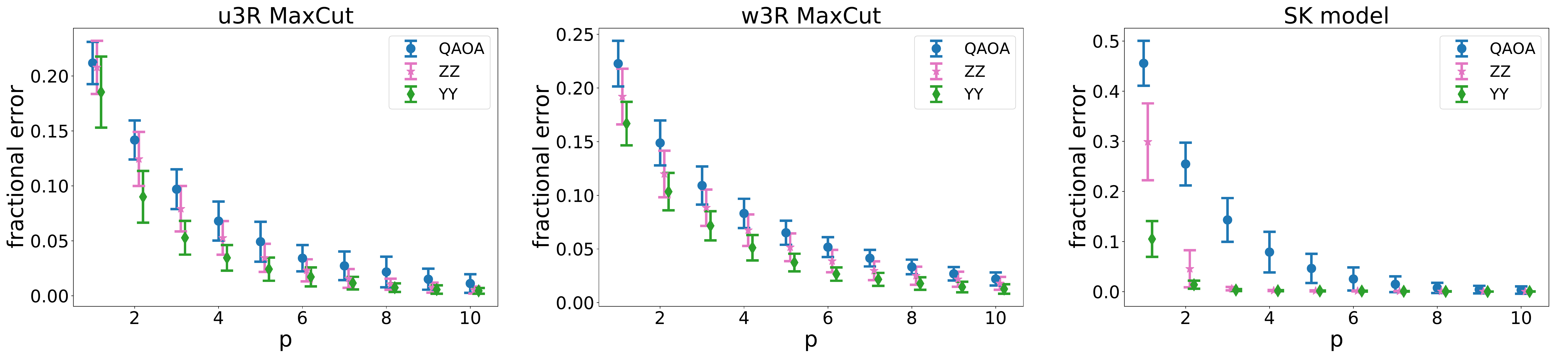}
\caption{Fractional errors with error bars. The results of each problem averaged over 20 random graphs, and the error bars are the standard deviation of the fractional error r for those random graphs.}
\label{fig:fig_error_bars}
\end{figure*}

The results in Fig.\ref{fig:u3r},\ref{fig:w3r},\ref{fig:sk} are averaged over the random graphs. Typically, the fractional error $r$ got by QAOA should be concentrated for the same class of problems. So it is better to show the results with error bars to represent the variance in the fractional error, and determine whether the performance difference between S-QAOA and QAOA is outside the margin of errors. Fig.\ref{fig:fig_error_bars} shows the fractional errors with error bars for Maxcut problems and SK model. Though there is a relatively large error bar because of the limited statistics, the performance of S-QAOA (labeled by `YY') is better than QAOA significantly.

\section{Additional Figures}

\begin{figure*}[htb]
	\centering
	\includegraphics[width=17cm]{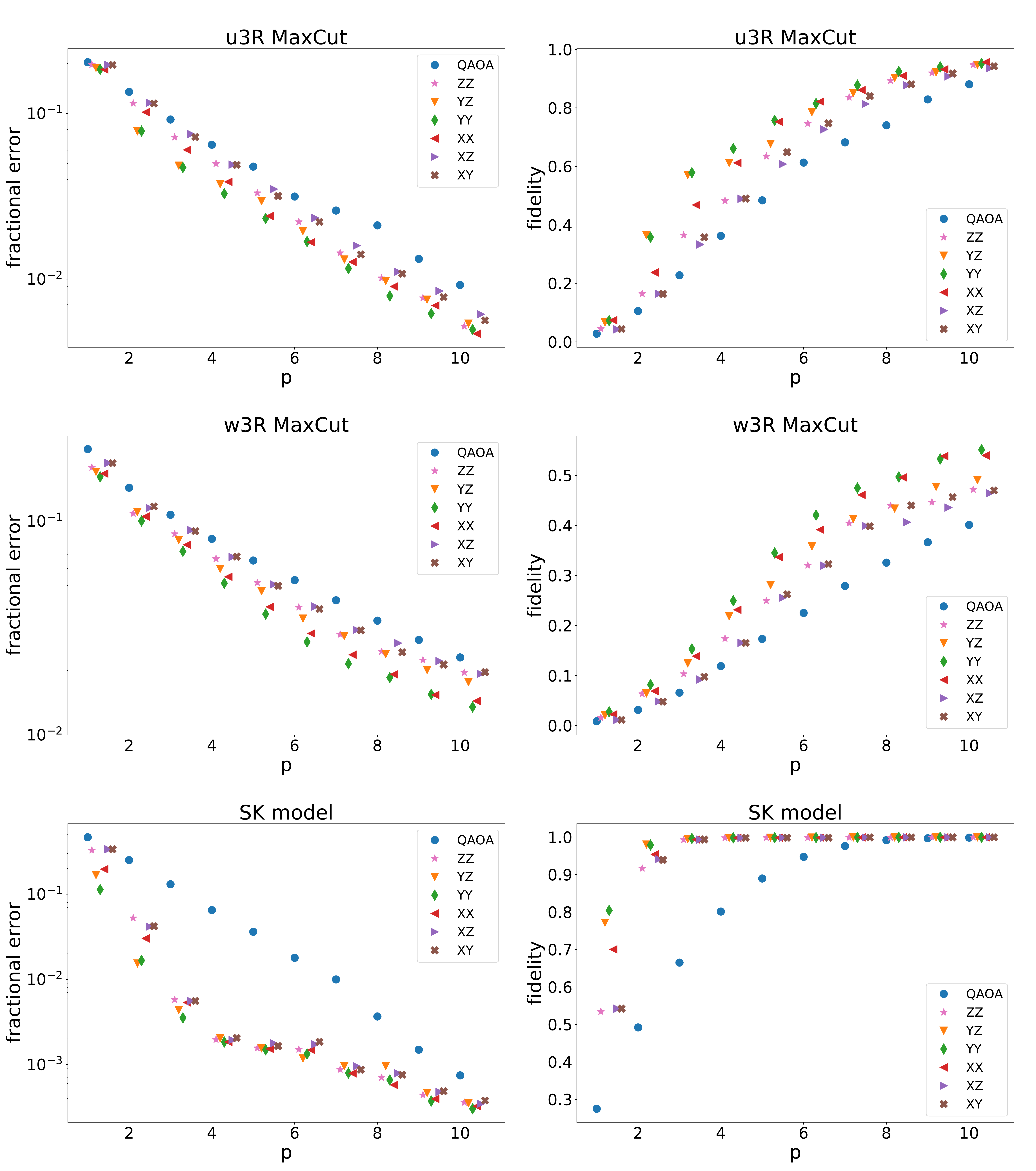}
	\caption{Comparison of the performance of all the possible additional two-body interactions: $\{YZ, YY, XX, XZ, XY\}$, the results are averaged on 10 random graphs for every case. YY interaction performs best for all the three cases we study. The performance of XZ, XY, and ZZ types are almost the same, so it demonstrates that the XZ and XY interactions have little effect on the QAOA.} 
	\label{fig:all_s_exp_poss}
\end{figure*}

\begin{figure*}[htb]
	\centering
	\includegraphics[width=17cm]{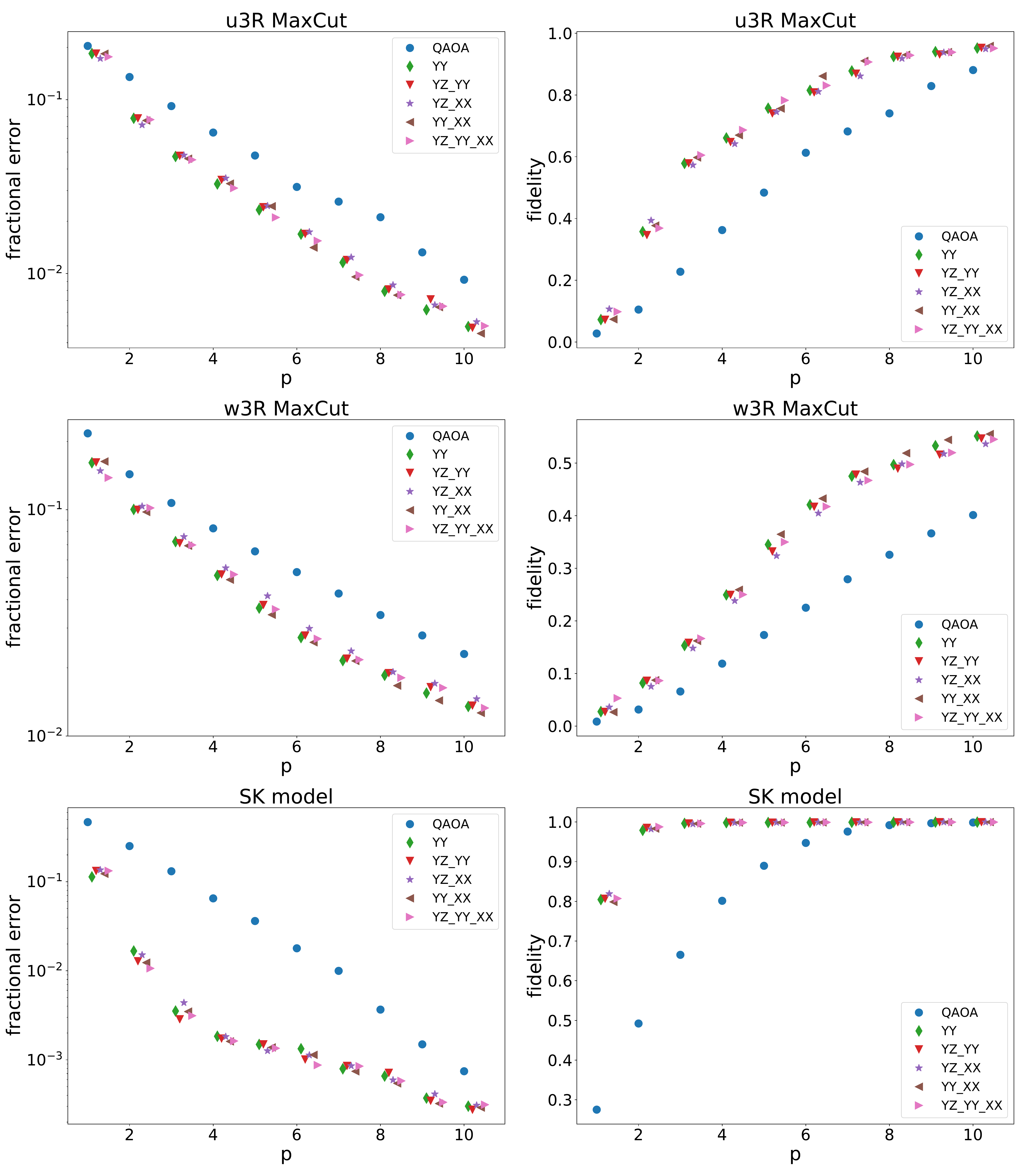}
	\caption{Comparison of the performance of only adding the YY interaction to the ansatz and adding two or three additional interactions to the ansatz, which are combination of: $\{YZ, YY, XX\}$ interactions, e.g., `YZ\_YY', `YZ\_XX', `YY\_XX' represent there are two additional interactions followed by each ZZ interaction; `YZ\_YY\_XX' represents the YZ, YY, XX interactions followed by ZZ interaction one by one. The results are averaged on 10 random graphs, and the performance of the ansatz that includes two or three additional interactions is almost the same as the ansatz that only includes the YY interaction. So if we consider two-gate terms, it is enough to add the YY interaction to accelerate the evolution.} 
	\label{fig:all_t_exp_poss}
\end{figure*}

Except for the existing ZZ interaction in QAOA ansatz, S-QAOA contains other two-body interactions to accelerate the process of quantum optimization. Operation of each layer $k \in [1,p]$ of S-QAOA is:
\begin{equation}
    \prod\limits_{k=1}^p\{e^{i\beta_k \sum\limits_i^n X_i} \prod\limits_{ij}\{e^{-i \gamma^{ij}_k w_{ij} Z_i Z_j} e^{-i\alpha_k \gamma^{ij}_k w_{ij} M_{ij}} \}\},
\end{equation}
$M_{ij}$ is a two-gate term and there are five possible types of it: $M_{ij} = \frac{P_iQ_j + Q_iP_j}{2}, PQ \in \{YZ, YY,XX,XZ,XY\}$. To choose the best one in these two-gate terms, we do a comprehensive simulation for MaxCut problem and SK model, and the result can be found in Fig.\ref{fig:all_s_exp_poss}. The simulation result implies that the $M_{ij} = Y_i Y_j$ has the best performance in all cases. A further study in Fig.\ref{fig:all_t_exp_poss} is to explore the necessity of adding more two-body interactions in the ansatz, and it is obvious that the ansatz adding more interactions can not get better performance than the ansatz just adding YY interaction. So it is enough to just add the YY interaction in S-QAOA ansatz.

%\nocite{*}

\end{document}